\documentclass[aps,prd,twocolumn,superscriptaddress]{revtex4}

\usepackage{graphicx}
\usepackage{amsmath}
\usepackage{ifthen}
\usepackage{subfigure}

\begingroup

\endgroup

\newcommand{\RotM}{\mathcal{R}}
\newcommand{\ktr}{\tilde{\kappa}_{\rm tr}}
\newcommand{\kem}{\tilde{\kappa}_{e-}}

\newcommand{\kop}{\tilde{\kappa}_{o+}}

\newcommand{\bx}{\beta_{x}}
\newcommand{\by}{\beta_{y}}
\newcommand{\bz}{\beta_{z}}

\newcommand{\omT}{\omega_{\oplus}T_{\oplus}}

\newcommand{\bop}{\beta_{\oplus}}

\newcommand{\epv}[2]{\ifthenelse{\equal{#2}{-k}}{\ensuremath{\vec{\epsilon}_{#1}(-\vec{k})}}{\ensuremath{\vec{\epsilon}_{#1}(\vec{#2})}}}

\newcommand{\kemxy}{\kem^{xy}}
\newcommand{\kemxz}{\kem^{xz}}

\newcommand{\kemyy}{\kem^{yy}}
\newcommand{\kemyz}{\kem^{yz}}

\newcommand{\kemzz}{\kem^{zz}}
\newcommand{\kopxy}{\kop^{xy}}
\newcommand{\kopyz}{\kop^{yz}}
\newcommand{\kopxz}{\kop^{xz}}

\newcommand{\kemxyp}{\kem'^{xy}}
\newcommand{\kemxzp}{\kem'^{xz}}

\newcommand{\kemyyp}{\kem'^{yy}}
\newcommand{\kemyzp}{\kem'^{yz}}

\newcommand{\kemzzp}{\kem'^{zz}}
\newcommand{\kopxyp}{\kop'^{xy}}
\newcommand{\kopyzp}{\kop'^{yz}}
\newcommand{\kopxzp}{\kop'^{xz}}

\begin{document}


\title{Improved Constraints on Isotropic Shift and Anisotropies of the Speed of Light using Rotating Cryogenic Sapphire Oscillators}


\author{Michael A. Hohensee}
\email{hohensee@berkeley.edu}
\affiliation{Department of Physics, Harvard University}
\affiliation{Department of Physics, University of California, Berkeley}
\author{Paul L. Stanwix}
\affiliation{School of Physics, The University of Western Australia}
\affiliation{Harvard-Smithsonian Center for Astrophysics}
\author{Michael E. Tobar}
\affiliation{School of Physics, The University of Western Australia}
\author{Stephen R. Parker}
\affiliation{School of Physics, The University of Western Australia}
\author{David F. Phillips}
\affiliation{Harvard-Smithsonian Center for Astrophysics}
\author{Ronald L. Walsworth}
\affiliation{Department of Physics, Harvard University}
\affiliation{Harvard-Smithsonian Center for Astrophysics}


\date{\today}

\begin{abstract}

We demonstrate that Michelson-Morley tests, which detect direction-dependent anisotropies in the speed of light, can also be used to place limits upon isotropic deviations of the vacuum speed of light from $c$, as described by the photon sector Standard Model Extension (SME) parameter $\tilde{\kappa}_{tr}$.  A shift in the speed of light that is isotropic in one inertial frame implies anisotropic shifts in others.  Using observer Lorentz covariance, we derive the time-dependent variations in the relative resonance frequencies of a pair of electromagnetic resonators that would be generated by such a shift in the rest frame of the Sun.  A new analysis of a recent experimental test of relativity using this result constrains $\tilde{\kappa}_{tr}$ with a precision of $7.4\times10^{-9}$. This represents the first constraint on $\tilde{\kappa}_{tr}$ by a Michelson-Morley experiment and the first analysis of a single experiment to simultaneously set limits on all nine non-birefringent terms in the photon sector of the SME.
\end{abstract}


\maketitle

\section{Introduction\label{sec:intro}}

Lorentz invariance is a cornerstone of both General Relativity and the Standard Model of Particle Physics, and as such has been the subject of many experimental investigations over the past century.  Much of this work has focused upon the properties and propagation of light in different reference frames, beginning with the pioneering work of Michelson-Morley~\cite{Michelson:1881}, Kennedy-Thorndike~\cite{Kennedy:1932}, and Ives-Stilwell~\cite{Ives:1938}. The purpose and interpretation of these experiments has varied with the development of physical theories throughout the century, ranging from attempts to observe the properties of a luminiferous aether, to determining whether space-time exhibits Lorentz as opposed to some other symmetry, and to more recent searches for the imprint of physics beyond the Standard Model.  These most recent studies presume that physics is invariant under ``passive'' transformations of the observer reference frame, while leaving open the possibility that the theory is not Lorentz invariant under ``active'' boosts of the rest frame of the system under test.
This could happen if known particles interact with fields not accounted for by the Standard Model, or indeed if Lorentz symmetry turns out to be explicitly broken.  In this context, modern implementations of Michelson-Morley, Kennedy-Thorndike, and Ives-Stilwell tests ~\cite{Stanwix:2006,Reinhardt:2007,Tobar:2009, Eisele:2009,Herrmann:2009} are used to look for evidence of such Lorentz violation in the form of modifications of the dispersion relation for light and other Standard Model particles.  In particular, these tests search for deviations of the phase velocity of light in vacuum from the canonical value.  These deviations can be orientation- and also polarization-dependent, and in general give the vacuum the properties of a potentially birefringent or anisotropic polarizable medium.  Such effects can be parameterized by the Standard Model Extension (SME)~\cite{Colladay:1997and1998, Kostelecky:2002}, which provides an effective field theory framework for determining the experimental consequences of a perturbative Lorentz violation. Observation of Lorentz violation in a physical system would provide clues about the structure of physics at experimentally inaccessible energy scales.

Modern Michelson-Morley experiments usually consist of a pair of orthogonally mounted electromagnetic resonators that are rotated in order to modulate their orientation in space. The observable is the difference in their resonant frequencies; Lorentz violations will manifest as periodic variations in the signal at frequencies related to the rotation and its harmonics.  Hence, such experiments are typically considered to be sensitive only to anisotropies in the speed of light.  Here, we extend the analysis of~\cite{Kostelecky:2002} to explicitly derive the sensitivity of Michelson-Morley tests to deviations in the speed of light that are isotropic in a given inertial reference frame.  Furthermore, using this result we report upon a new analysis of data from a recent experiment ~\cite{Stanwix:2006} that constrains all nine non-birefringent CPT-even photon-sector SME coefficients, summarized in table \ref{tab:results}. In particular, our analysis constrains the isotropic shift parameter $|\ktr|$, the first such result from this form of experiment.  
Although this constraint is overshadowed by recent results based on collider physics~\cite{Altschul:2009b,Hohensee:2009}, it is an improvement upon results obtained from experiments intended to constrain $\ktr$ such as relativistic ion spectroscopy ~\cite{Reinhardt:2007}.  

\begin{table}[t!]
\caption[Results]{Fitted values and uncertainties of the non-birefringent photon-sector parameters of the SME for the results reported here.  ($\kem$ in $10^{-16}$, $\kop$ in $10^{-12}$ and $\ktr$ in $10^{-8}$).}
\begin{ruledtabular}
\begin{tabular}{l r | l r | l r}
$\kem^{XY}$ & 0.8 (0.6) & $\kem^{XX}-\kem^{YY}$ & 0.2 (1.0) & $\kop^{XY}$ & -1.5 (1.2)  \\
$\kem^{XZ}$ & 1.5 (1.3) & $\kem^{ZZ}$ & 143 (179) & $\kop^{XZ}$ & 1.7 (0.7)  \\
$\kem^{YZ}$ & 1.7 (1.3) & $\ktr$ & -1.5 (0.74) & $\kop^{YZ}$ & 0.2 (0.7)  
\end{tabular} 
\end{ruledtabular}
\label{tab:results}
\end{table}

\section{Michelson-Morley Tests of the SME\label{sec:resbound}}

In general, Lorentz violation in the electromagnetic sector of the SME causes vacuum birefringence and polarization-independent shifts in the phase velocity of light in vacuum ($c_{\rm ph}$) relative to the canonical velocity ($c$). Vacuum birefringence has been constrained to better than one part in $10^{37}$ by observations of linearly polarized light from distance gamma ray bursts~\cite{Kostelecky:2006}, so is neglected in this analysis. 
The remaining polarization-independent shifts can be parameterized for a specified reference frame (e.g., the frame in which the sun is at rest) using nine degrees of freedom: one to describe the average deviation of $c_{\rm ph}$ from $c$ over all possible directions of propagation, five to describe the difference in the average speed of light moving forward and backwards along any given direction, and three more to describe the difference in $c_{\rm ph}$ for light moving in one direction relative to a counterpropagating beam.  To leading order, the SME uses the scalar $\ktr$, the $3\times 3$ symmetric traceless $\kem^{jk}$ matrix with five degrees of freedom, and the $3\times 3$ antisymmetric $\kop^{jk}$ matrix with three degrees of freedom to parameterize these shifts.  In terms of these $\tilde{\kappa}$'s, the free electromagnetic Lagrangian becomes~\cite{Kostelecky:2002}
\begin{multline}
\label{lagr2}
\!\!\!\!\!
{\cal L}=\frac{1}{2}\left[(1+\ktr)\vec{E}^{2}-(1-\ktr)\vec{B}^{2}\right]+\vec{E}\cdot(\kop)\cdot\vec{B}\\+\frac{1}{2}\vec{E}\cdot(\kem)\cdot\vec{E}
+\frac{1}{2}\vec{B}\cdot(\kem)\cdot\vec{B}\;,
\end{multline}
where $\vec{E}$ and $\vec{B}$ are the standard electromagnetic fields in vacuum.

Although the total Lagrangian remains invariant under changes in an observer's inertial frame, the parts proportional to the $\tilde{\kappa}$ coefficients are not term by term invariant.  If, for example, the speed of light in a reference frame $S$ is $c_{\rm ph+}$ for a wave with wavevector $\vec{k}$, and $c_{\rm ph-}$ for waves traveling in the opposite direction, and $c_{\rm ph+}=c_{\rm ph-}\neq c$, then observer Lorentz invariance requires that the phase velocity of these two waves must differ from one another in any reference frame $S'$, arrived at from $S$ via a boost along $\vec{k}$.  This difference must also be reflected in the values taken by the $\tilde{\kappa}$'s when the Lagrangian is expressed in terms of the fields in $S'$.
The $\tilde{\kappa}$'s mix with one another under rotations and boosts of the observer coordinate frame. Therefore, results from a series of identical experiments performed in different inertial frames may be used to obtain constraints on all nine of the non-birefringent $\tilde{\kappa}$'s, even though any individual experiment might only be sensitive to a subset.

It is convenient to select a standard inertial frame in which to compare the results of different experimental tests of Lorentz invariance, and to express the numerical values (or limits) on the SME coefficients.  We adopt the Sun Centered Celestial Equatorial Frame (SCCEF), following~\cite{Kostelecky:2002,Kostelecky:2008z}, which is defined with the coordinate origin at the Sun, X and Y lie in the plane of the Earth's equatorial plane, with X pointing towards the Earth at the autumnal equinox.

Let us now consider the Michelson-Morley laboratory experiment. Following on from Eq. \eqref{lagr2}, it can be shown that if any of the $\tilde{\kappa}$ parameters are nonzero the difference frequency between the electromagnetic modes of a pair of identical, orthogonally mounted resonators is given by~\cite{Kostelecky:2002,Tobar:2006}
\begin{equation}
\frac{\delta \nu}{\nu} = S_{e}\left\{\left[(\kem)^{xx}_{\rm lab}-(\kem)^{yy}_{\rm lab}\right]\cos {2\theta}-2(\kem)^{xy}_{\rm lab}\sin{2\theta}\right\},\label{eq:michelsonmorleysens}
\end{equation}
where $S_{e}$ is a sensitivity factor specific to the resonator modes and materials, $\theta$ is the angle of the resonators' axes relative to the $x$ and $y$ coordinate axes, which are in turn defined by the system configuration when $\theta=0$.  Thus, in a given inertial reference frame Michelson-Morley experiments directly constrain the value of $\kem^{jk}$ in the laboratory. In practice, however, changes in the Earth's motion relative to the Sun during its orbit also allow us to set limits on the magnitudes of $\kop^{jk}$~\cite{Kostelecky:2002} in the Sun's rest frame. We show in this work that the relationship can be further extended to constrain the magnitude of $\ktr$, as is derived in detail in Appendix~\ref{app:allorder}, and outlined in the following section.

\section{Sensitivity to the isotropic $\ktr$\label{sec:sens}}
{
\begin{table*}[ht!]
\caption[Contribution of $\ktr$ to Michelson-Morley Experiments]{Contributions of $\ktr$, as defined in the SCCEF, to the amplitude of sidereal variations in the Michelson-Morley observable normalized for the experimental sensitivity $S_{e}$, in terms of the relative orientation and boost of the laboratory frame relative to the SCCEF.  $\eta$ is the declination of the Earth's orbit relative to its spin, taken to be $23.27^{\circ}$, and $\chi$ is the colatitude of the laboratory, $121.82^{\circ}$ in Perth, Australia.  The actual magnitude of each signal due to $\ktr$ for an experiment in Perth, Australia is indicated by the numerical weight.  Although $\ktr$ does generate signals at the frequency $\omega_{\oplus}$ of the sidereal day, and also at $2\omega_{\oplus}$, the magnitude of such contributions is strongly suppressed relative to those from $\kem$ and $\kop$, which respectively are of order unity and $10^{-4}$.  At all other frequencies, the signals from $\kem$ and $\kop$ are suppressed relative to $\ktr$.  This, combined with the far more stringent bounds set upon $\kem$ and $\kop$ from other experiments \cite{Mueller:2007}, allows us to ignore all but the contribution of $\ktr$ to signals at $\omega_{\oplus}\pm2\Omega_{\oplus}$ and $2\omega_{\oplus}\pm2\Omega_{\oplus}$, where $\Omega_{\oplus}$ is the frequency of the sidereal year.}
\begin{tabular}{c|cccr}
\hline\hline
$\omega_{i}$ & $C_{C,\omega_{i}}$ & \parbox{2cm}{Num. Weight ($\times 10^{-10}$)} & $C_{S,\omega_{i}}$ &  \parbox{2cm}{Num. Weight ($\times 10^{-10}$)}  \\
\hline
$\omega_{\oplus}$ & - & & $\tfrac{1}{2}\bop^{2}\sin{2\eta}\sin{2\chi}\ktr$ & -32.1 \\
$2\omega_{\oplus}$ & $\tfrac{-1}{2}\bop^{2}\sin^{2}{\eta}(1+\cos^{2}{\chi})\ktr$ & -9.85 & -\\
$\omega_{\oplus}+2\Omega_{\oplus}$ & - & & $\tfrac{-1}{2}\bop^{2}\sin{2\chi}(1-\cos{\eta})\sin{\eta}\ktr$ & $1.42$ \\
$\omega_{\oplus}-2\Omega_{\oplus}$ & - & & $\tfrac{1}{2}\bop^{2}\sin{2\chi}(1+\cos{\eta})\sin{\eta}\ktr$ & $-33.5$\\
$2\omega_{\oplus}+2\Omega_{\oplus}$ & $\tfrac{1}{4}\bop^{2}(1+\cos^{2}{\chi})(1-\cos{\eta})^{2}\ktr$ & $0.209$ & - \\
$2\omega_{\oplus}-2\Omega_{\oplus}$ & $\tfrac{1}{4}\bop^{2}(1+\cos^{2}{\chi})(1+\cos{\eta})^{2}\ktr$ & $116$ & - \\
\hline\hline
$\omega_{i}$ & $S_{C,\omega_{i}}$ & \parbox{2cm}{Num. Weight ($\times 10^{-10}$)} & $S_{S,\omega_{i}}$ &  \parbox{2cm}{Num. Weight ($\times 10^{-10}$)} \\
\hline
$\omega_{\oplus}$ & $-\bop^{2}\cos{2\eta}\sin{\chi}\ktr$ & -60.9 & - & \\
$2\omega_{\oplus}$ & - & & $-\bop^{2}\sin^{2}{\eta}\cos{\chi}\ktr$ & 8.13\\
$\omega_{\oplus}+2\Omega_{\oplus}$ & $\bop^{2}\sin{\chi}\sin{\eta}(1-\cos{\eta})\ktr$ & $6.82$ & - \\
$\omega_{\oplus}-2\Omega_{\oplus}$ & $-\bop^{2}\sin{\chi}\sin{\eta}(1+\cos{\eta})\ktr$ & $-161$ & - \\
$2\omega_{\oplus}+2\Omega_{\oplus}$ & - & & $\tfrac{1}{2}\bop^{2}\cos{\chi}(1-\cos{\eta})^{2}\ktr$ & $-0.172$\\
$2\omega_{\oplus}-2\Omega_{\oplus}$ & - & & $\tfrac{1}{2}\bop^{2}\cos{\chi}(1+\cos{\eta})^{2}\ktr$ & $-95.9$\\ 
\hline\hline
\end{tabular} 
\label{tab:coef}
\end{table*}
}Although the general form of the time-dependence of the Earthbound lab-frame $\tilde{\kappa}$'s can be derived using the observer covariance of the action, the terms contributing to \eqref{eq:michelsonmorleysens} proportional to the value of $\ktr$ in the Sun-Centered Celestial Equatorial Frame (SCCEF) can be obtained using simpler arguments.  The resonator in a Michelson-Morley experiment is sensitive only to anisotropies that the SCCEF $\ktr$ generates in the laboratory frame, which in turn must depend solely upon the orientation of the lab with respect to the lab's boost relative to the SCCEF.  The maximum difference signal is generated when the axis of one resonator is most parallel to the boost from the SCCEF, while the axis of the other is as nearly perpendicular to the boost as possible.  In general, this will happen twice per solar day, although the precise times that they occur will vary over the course of a year.  For example, an experiment with one cavity axis aligned East to West in the lab sees a peak daily $\ktr$-induced shift in that cavity maximized during the summer and winter solstices, while its peak shift is minimized at the equinoxes.  

Since the frequency shift between two identical resonators given by Eq. \eqref{eq:michelsonmorleysens} is the same up to a constant factor for any such pair~\cite{Tobar:2006}, we can analyze the simple case of a pair of Fabry-Perot cavities aligned orthogonally to one another along the $x$- and $y$-axes.  The resonance frequencies of each of the cavities are then $\nu=\frac{m c_{\rm ph}}{2L}$, where $L$ is the length of the cavity, $m=1,2,3,\dots$, and $c_{\rm ph}=(c_{\rm ph}^{+}+c_{\rm ph}^{-})/2$ is the average phase velocity of light moving back and forth along the cavities' axes.  Variations in the phase velocity of light along the cavity axes yields the frequency difference
\begin{equation}
\frac{\delta\nu_{x}}{\nu_{x}}-\frac{\delta\nu_{y}}{\nu_{y}}=\frac{1}{2}(\rho_{x+}+\rho_{x-}-\rho_{y+}-\rho_{y-}),\label{eq:freqdiff}
\end{equation}
where $c\rho_{j\pm}=\delta c_{j\pm}$ is the shift in the vacuum phase velocity of light parallel ($+$) or anti-parallel ($-$) to the $j$-axis due to $\tilde{\kappa}$ in the laboratory frame.  The problem now reduces to finding the mean speed of light along the laboratory $x$- and $y$-axes in terms of $\ktr$ in the SCCEF.  

Observer Lorentz covariance of the SME implies that the overall Lagrangian remains a Lorentz scalar, although the action may not be term-by-term Lorentz invariant~\cite{Colladay:1997and1998}.  This means that although the speed of light might be frame-dependent, the velocity of a particular electromagnetic wave must transform in the same manner as any other velocity under Lorentz boosts --i.e. according to the relativistic velocity addition formula.  Given the boost of the laboratory relative to the SCCEF, we can then determine the anisotropic shift in the speed of light as seen in the laboratory arising from an isotropic shift (nonzero $\ktr$) in the SCCEF.  We seek a solution that is leading order in the $\tilde{\kappa}$'s, and so the laboratory anisotropies induced by a nonzero $\ktr$ in the SCCEF can depend only upon $\ktr$.  To second order in the laboratory boost $\beta=v/c$ relative to the SCCEF, we then obtain
\begin{equation}
\frac{1}{2}(\rho_{x+}+\rho_{x-})=-\ktr-(\beta^{2}+\beta_{x}^{2})\ktr,
\end{equation}
and similarly for the mean speed of light along the $y$-axis.  More details of this derivation can be found in Appendix \ref{app:relveladd}.  The differential signal produced by a pair of orthogonally mounted resonators must then be given by
\begin{equation}
\frac{\delta\nu}{\nu}\propto\frac{\delta\nu_{x}}{\nu_{x}}-\frac{\delta\nu_{y}}{\nu_{y}}=(\beta_{y}^{2}-\beta_{x}^{2})\ktr.\label{eq:secondordermmktrdep}
\end{equation}

For an experiment which rotates about the laboratory $z$-axis with angular frequency $\omega_{R}$, we find that the variation of the Lorentz-violating frequency shift in time is given by
\begin{equation}
\frac{\delta\nu}{\nu}=S(T)\sin{2\omega_{R}T}+C(T)\cos{2\omega_{R}T},\label{eq:timedepsig}
\end{equation}
where
\begin{align}
S(T)&=(\ktr)\times\sum_{i}\Big[S_{S,i}\sin(\omega_{i}T)+S_{C,i}\cos(\omega_{i}T)\Big],\label{eq:S}\\
C(T)&=(\ktr)\times\sum_{i}\Big[C_{S,i}\sin(\omega_{i}T)+C_{C,i}\cos(\omega_{i}T)\Big].\label{eq:C}
\end{align}
The overall modulation of the signal by $2\omega_{R}$ follows from the fact that \eqref{eq:freqdiff} is unchanged when we exchange $+x\leftrightarrow -x$ and $+y\leftrightarrow -y$.  The remaining modulation frequencies $\omega_{i}$ are various harmonics of and beats between the frequencies of the sidereal day $\omega_{\oplus}$ and the sidereal year $\Omega_{\oplus}$, as derived in the Appendicies.  
  The weights $S_{S,i}$, $S_{C,i}$, $C_{S,i}$, and $C_{C,i}$ most relevant to $\ktr$ are summarized in table \ref{tab:coef}.

\section{Experiment and Analysis\label{sec:exp}}

The bounds presented here arise out of a new analysis of data from an experiment performed at the University of Western Australia \cite{Stanwix:2006}. This experiment searched for Lorentz violating signals by monitoring the difference frequency between two microwave cryogenic sapphire oscillators (CSOs) as a function of orientation and time. The details of this experiment and the operation of CSOs in general has been reported elsewhere \cite{Locke:2008, Stanwix:2005, Stanwix:2006, Tobar:2006}, so we will provide only a brief description here. Each CSO relies upon a high Q-factor ($\sim 2\times10^{8}$) sapphire loaded cylindrical resonant cavity, excited in the $WGH_{8,0,0}$ whispering gallery mode at approximately 10 GHz by a Pound stabilized loop oscillator circuit.  The two resonators are mounted one above the other with their cylindrical axes orthogonal in the horizontal plane.  The experiment was continuously rotated in the laboratory with a period of 18 seconds.  When resonantly excited, the sapphire crystals support standing waves with the dominant electric and magnetic fields pointing in the axial and radial directions respectively.  For such whispering gallery modes, the Poynting vector is directed around the crystal circumference.  The resonant frequency of each crystal is directly proportional to the integrated phase velocity of light along the closed path followed by the resonant mode, and is thus sensitive to Lorentz violation in the photon sector of the SME.  Note that in this experiment a significant fraction of the mode field exists in the sapphire crystal, therefore the resonance frequency could also be perturbed by Lorentz violation in the electron sector. However, the relevant SME parameters for electrons have been constrained by other experiments \cite{Altschul:2006, Mueller:2007, Kostelecky:2008z} to the degree that they do not make significant contributions to these results, so we assume that electrons are fully Lorentz-symmetric.  

\begin{figure}[h]
\centering
\subfigure[]
{
\includegraphics[width=3in]{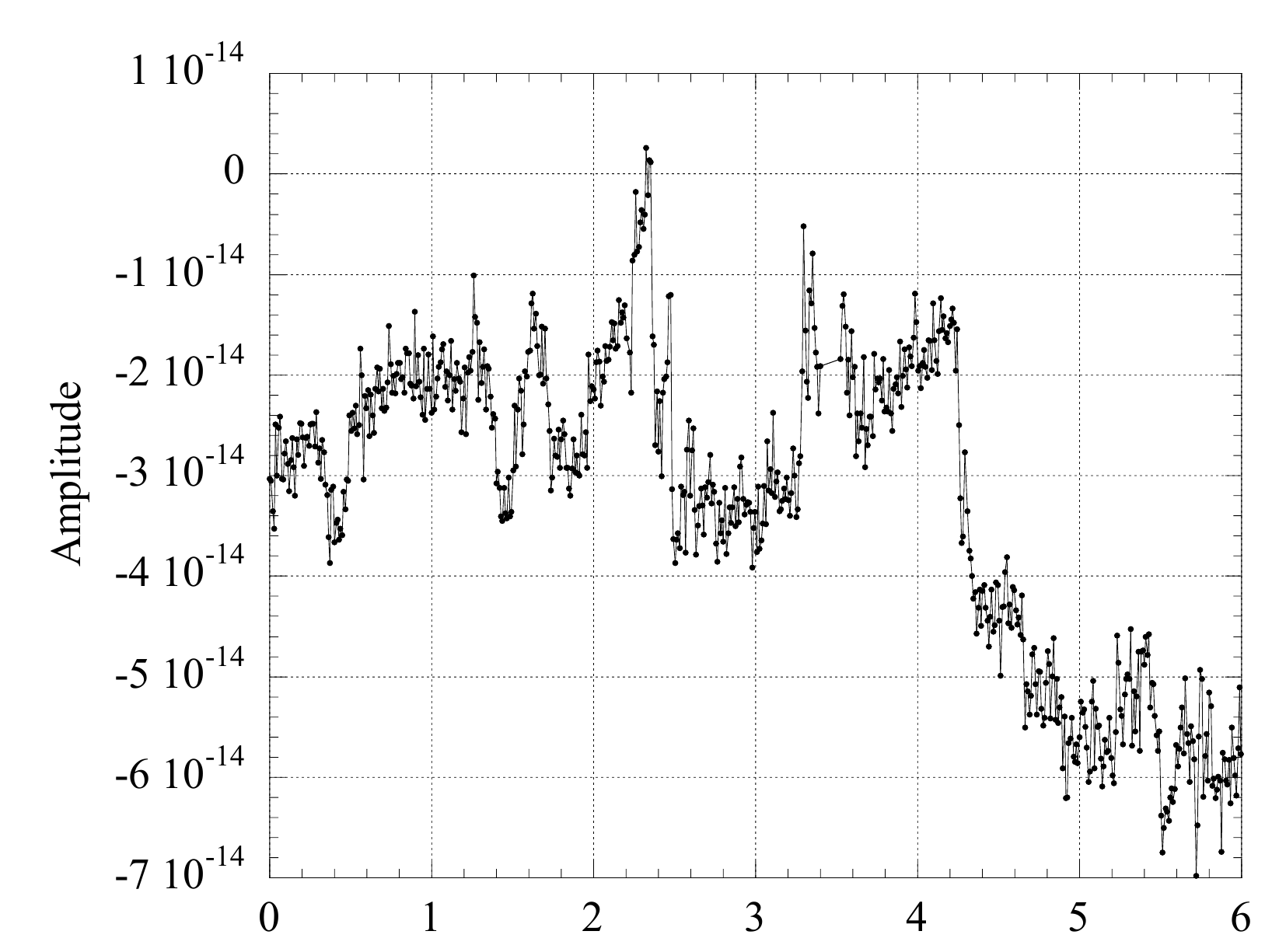} 
\label{fig:rawdat_a}
}\\
\subfigure[]
{
\includegraphics[width=3in]{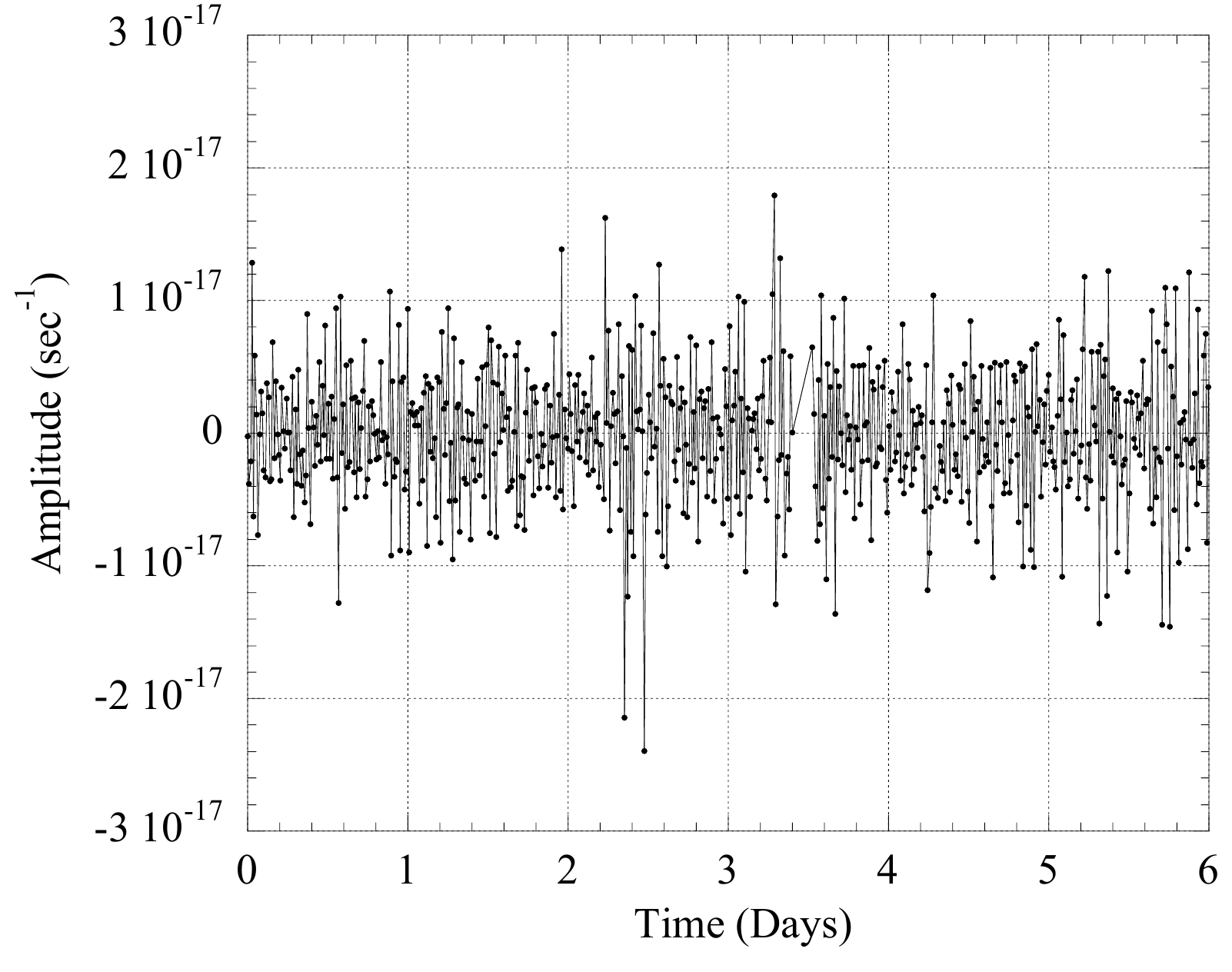}
\label{fig:rawdat_b}
}
\caption[June 17th S(T) and Derivative]{(a) Amplitudes of S(T) and (b) $\frac{dS(T)}{dT}$ obtained by demodulating the data at $2\omega_{R}$ in blocks of 50 rotations. Data was collected from the 17th - 23rd of June 2005. \label{fig:rawdat}}
\end{figure}

Data was collected from this experiment over a period of 400 days, with a useful duty cycle of 30$\%$. The data analysis is complicated by three main issues, each of which are addressed using specific techniques that in turn constitute the 3 steps of our analysis process. The first is the size of the data set. Processing the entire data set simultaneously is computationally intensive, so the data is initially reduced using the same technique described in \cite{Stanwix:2006}. The data is demodulated in quadrature at twice the frequency of the experiment rotation over an integer number of cavity rotation periods, $m$, generating a reduced demodulated data set consisting of S(T$_{i}$) and C(T$_{i}$) coefficients of equations \eqref{eq:S} and \eqref{eq:C}, centered at the mean time of the demodulated data block, T$_{i}$. This reduces the size of the data set by $12\times m$ (12 measurements during each of the $m$ rotations). Figure \ref{fig:rawdat_a} shows a typical subset of the data acquired continuously over 6 days, demodulated in blocks of 50 periods. In addition to reducing the size of the data, demodulation also effectively filters noise. In the final analysis for the results presented here we chose to use 500 periods, which maximizes the signal to noise ratio of the data while satisfying the Nyquist sampling rate (providing more than 2 data points per half day). 

The second main issue is the presence of jumps in the data, which are due to non-stationary noise sources such as sudden stress release in the resonator \cite{TobarIvanov:2006}. When analyzing the data using standard regression techniques, such as Least Squares, these jumps mimic temporal signals resulting in incorrect parameter estimates. One solution to this problem is to remove short sections of data containing these jumps, identified using an unbiased method, albeit at the cost of reducing the useful duty cycle of the experiment. In this work we employed an alternative approach of taking the derivative of the data, which involves differencing successive data points~\cite{Tobar:2010}. Signal jumps manifest in the derivative as singular outliers (illustrated in Figure \ref{fig:rawdat_b}) to which the Least Squares analysis is less susceptible. This is preferable since no data is excluded, the signal to noise is maximised, and no bias is applied to the data. For nonzero $\tilde{\kappa}$'s, the derivative of the data will vary according to the derivative of \eqref{eq:S} and \eqref{eq:C}:
\begin{equation}
\frac{dS(T)}{dT}=(\ktr)\times\sum_{i}\Big[\omega_{i}S_{S,i}\cos(\omega_{i}T)-\omega_{i}S_{C,i}\sin(\omega_{i}T)\Big],\label{eq:dS}
\end{equation}
\begin{equation}
\frac{dC(T)}{dT}=(\ktr)\times\sum_{i}\Big[\omega_{i}C_{S,i}\cos(\omega_{i}T)-\omega_{i}C_{C,i}\sin(\omega_{i}T)\Big].\label{eq:dC}
\end{equation}

\begin{figure*}[ht]
\centering
\includegraphics[width=6.5in]{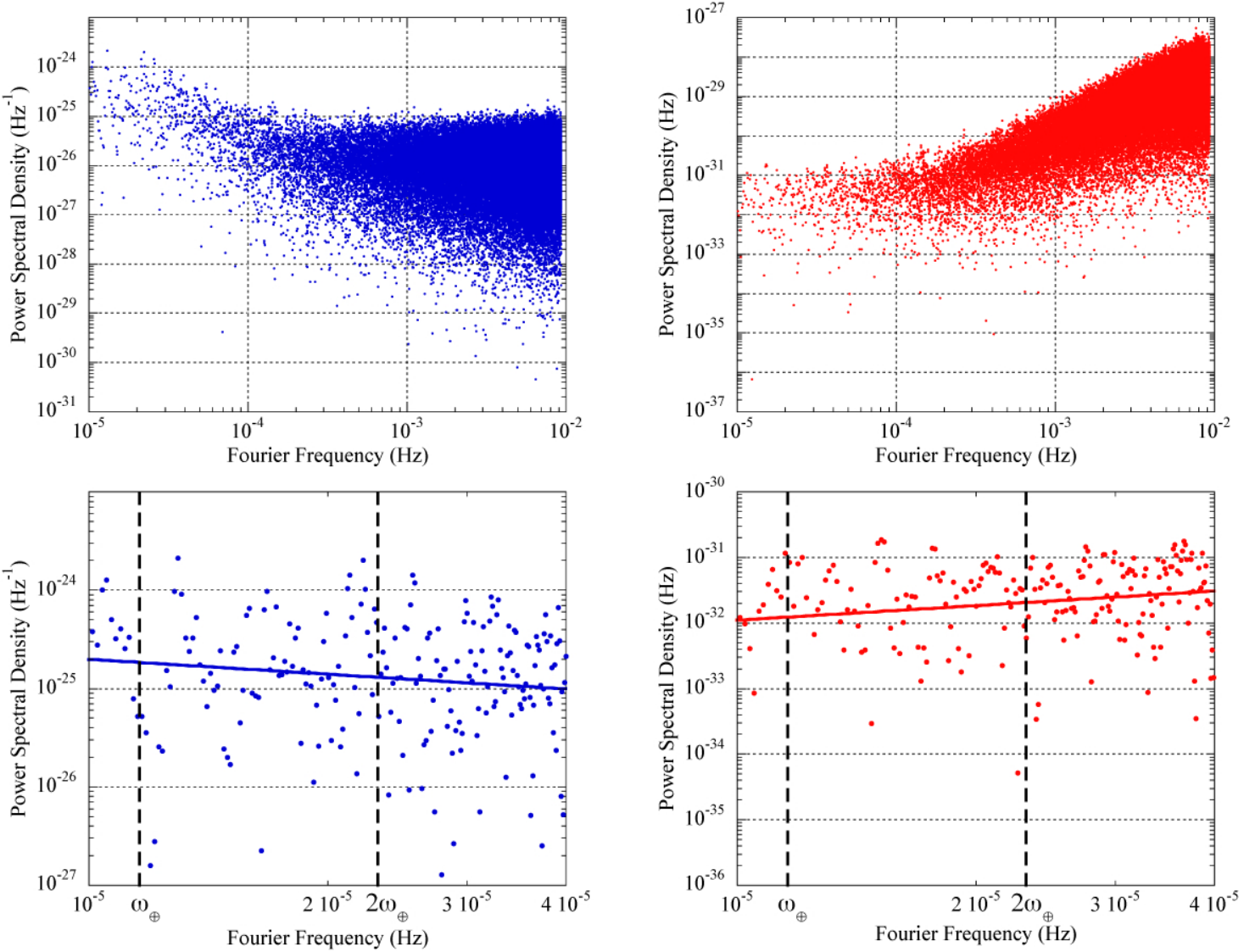}
\caption[Power Spectral Density of Residuals]{Power spectral densities of residuals from the S(T) demodulated (averaged over 2 rotations) least squares data analysis (top graphs) of the normal data (blue curves) and the derivative of the data (red curves). Power laws are fitted around the frequencies of interest (bottom graphs). \label{fig:psd}}
\end{figure*}

The third and final step of the analysis is to fit the frequencies of interest to the data using Least Squares regression.  Ordinary Least Squares (OLS) regression assumes that the Power Spectral Density (PSD) of the residuals is white.  Figure \ref{fig:psd} shows the PSD of the data following demodulation over 2 periods of rotation (Increasing the number of rotations over which the data is averaged truncates the PSD curves, acting as a low pass filter). For frequency offsets above $10^{-4}$ Hz the noise is white; near the frequencies of interest, $\omega_{\oplus}$ and $2\omega_{\oplus}$ ($\sim10^{-5}$ Hz), however, a power law with  $\alpha=0.5$ describes the power spectral density. Similarly, once differentiated, the PSD exhibits a power law with $\alpha=0.75$, which is then used in the third part of this analysis. To account for the noise color of the data we use a Weighted Least Squares (WLS) technique that whitens the noise by pre-multiplying the data and the fit model with a weighting matrix. The weighting matrix is determined using a fractional differencing technique \cite{Schmidt:2003} that corrects for serially correlated noise, as determined from $\alpha$.  Different frequencies are used to set limits on $\ktr$ and the $\kem$ and $\kop$ components, allowing a simultaneously fit of all nine components using the coefficients in table \ref{tab:coef} and the others already derived in \cite{Stanwix:2006}.

\section{Conclusion\label{sec:conclusion}}

Using a more sophisticated analysis of data collected in~\cite{Stanwix:2006}, we have tightened the limits set by this experiment on the magnitude of all the non-birefringent $\tilde{\kappa}$ coefficients of the SME by a factor between 1.5 and 4, as summarized in table \ref{tab:results}.  We have explicitly demonstrated that Michelson-Morley experiments are sensitive to isotropic shifts in the vacuum speed of light, and thus for the first time, we report a simultaneous set of bounds on all nine of the non-birefringent $\tilde{\kappa}$ coefficients.  The new limit on $\ktr$ is an improvement of more than a factor of 11 over limits obtained by relativistic ion spectroscopy~\cite{Reinhardt:2007}, marking the first time that a low energy experiment has been able to surpass the sensitivity of such tests.

%
\begin{acknowledgments}
This work was supported by the National Science Foundation and the Australian Research Council.  We thank Alan Kostelecky for encouragement and useful discussions.
\end{acknowledgments}

\appendix
\section{\label{app:relveladd}}

Consider a beam of light moving with velocity $\vec{u}$ along the $x$-axis in the laboratory, which itself moves with velocity $\vec{v}=c\vec{\beta}$ relative to the SCCEF, and define
\begin{align}
\vec{u}_{||}&=\frac{\vec{v}\cdot\vec{u}}{|v|^{2}}\vec{v} & \vec{u}_{\perp}&=\vec{u}-\vec{u}_{||}.
\end{align}
The velocity $\vec{s}$ of that beam of light as measured in the SCCEF must be
\begin{equation}
\vec{s}/c=\frac{\vec{v}/c+\vec{u}_{||}/c+\vec{u}_{\perp}/(c\gamma)}{1+\vec{v}\cdot\vec{u}/c^{2}}.\label{eq:veltrans}
\end{equation}
Since we are interested solely in the contribution of the SCCEF $\ktr$ to our experiment, and not in terms proportional to products of the $\tilde{\kappa}$'s, we may assume $\rho_{x\pm}$ and $\rho_{y\pm}$ are such that the speed of light in the SCCEF is isotropic and equal to $c(1-\ktr)$.  Taking the norm of \eqref{eq:veltrans} yields
\begin{equation}
(1-\ktr)^{2}=\frac{\beta^{2}-(1+\rho_{x\pm})\left((1+\rho_{x\pm})(\beta^{2}-\beta_{x}^{2}-1)\mp2\beta_{x}\right)}{\left(1\pm\beta_{x}(1+\rho_{x\pm})\right)^{2}},
\end{equation}
which to second order in $\beta$ and first order in $\ktr$, becomes
\begin{equation}
\frac{1}{2}\left(\rho_{x+}+\rho_{x-}\right)=-\ktr-(\beta^{2}+\beta_{x}^{2})\ktr.
\end{equation}
Note that we have neglected terms proportional to $\ktr\rho_{x\pm}$, since $\rho_{x\pm}$ is of the same order as $\ktr$.  We can repeat the above argument to obtain the mean velocity of light along the $y$-axis to find that the dependence of $\delta\nu/\nu$ on $\ktr$ is given by
\begin{equation}
\frac{\delta\nu}{\nu}\simeq\frac{\delta\nu_{x}}{\nu_{x}}-\frac{\delta\nu_{y}}{\nu_{y}}=\left(\beta_{y}^{2}-\beta_{x}^{2}\right)\ktr.\label{eq:secondordermmktrdepapp}
\end{equation}
The detailed form of the boost $\vec{\beta}$ from the SCCEF as defined in the laboratory frame is~\cite{Kostelecky:2002}
\begin{equation}
\vec{\beta}=\RotM\left(\begin{matrix}\beta_{\oplus}\sin\Omega_{\oplus}T\\ -\beta_{\oplus}\cos\eta\cos\Omega_{\oplus}T\\ -\beta_{\oplus}\sin\eta\cos\Omega_{\oplus}T\end{matrix}\right),
\end{equation}
where we have neglected the contribution of the earth's rotation $\beta_{L}\simeq 10^{-6}$ to the boost vector, $T$ is the time since the last vernal equinox, and the rotation $\RotM$ which reorients the SCCEF to align with the laboratory frame, with $\hat{z}$ pointing upwards and $\hat{x}$ pointing south, is given by
\begin{equation}
\RotM=\left(\begin{matrix} \cos\chi\cos\omT & \cos\chi \sin\omT & -\sin\chi\\ -\sin\omT & \cos\omT & 0\\ \sin\chi\cos\omT & \sin\chi\sin\omT & \cos\chi\end{matrix}\right).
\end{equation}
Here $\chi$ is the colatitude of the laboratory, $\eta$ is the declination of the Earth's orbit relative to its spin, $\omega_{\oplus}$ and $\Omega_{\oplus}$ are the Earth's annual and sidereal frequencies, and $\beta_{\oplus}\simeq 0.994\times 10^{-4}$ is the Earth's orbital speed.  The time $T_{\oplus}$ is not the same as $T$, and represents the time as measured in the SCCEF since that frame's $Y$-axis coincided with the laboratory $y$-axis~\cite{Kostelecky:2002}.  We can account for the active rotation of the experiment~\cite{Stanwix:2006} by redefining $\RotM$ so as to be aligned with the resonator axes:
\begin{multline}
\RotM=\left(\begin{matrix} \cos\omega_{R}T & -\sin\omega_{R}T & 0\\ \sin\omega_{R}T & \cos\omega_{R}T & 0 \\ 0 & 0 & 1\end{matrix}\right)\\
\cdot\left(\begin{matrix} \cos\chi\cos\omT & \cos\chi \sin\omT & -\sin\chi\\ -\sin\omT & \cos\omT & 0\\ \sin\chi\cos\omT & \sin\chi\sin\omT & \cos\chi\end{matrix}\right).
\end{multline}
Insertion of $\vec{\beta}$ into \eqref{eq:secondordermmktrdepapp} yields modulations with the form of Eq.~\eqref{eq:timedepsig}, described in part \ref{sec:sens}.

\section{\label{app:allorder}}

This appendix presents the general form of transformations of the non-birefringent $\tilde{\kappa}$ coefficients under an arbitrary boost $\vec{\beta}$ from one inertial frame ($S$) to another ($S'$).  This derivation rests upon the assumption of Lorentz invariance under observer transformations: that the Lagrangian is an overall Lorentz scalar quantity.  In what follows, we neglect the contribution of SME coefficients which give rise to vacuum birefringence, as these have been constrained to be of order $10^{-37}$ or less~\cite{Kostelecky:2006}, and do not contribute to the non-birefringent physics at leading order~\cite{Colladay:1997and1998}.  With this assumption in hand, we may begin with the Lagrangian of Eq.~\eqref{lagr2}, defined in frame $S$ in terms of the fields $\vec{E}$ and $\vec{B}$ as
\begin{multline}
\!\!\!\!\!
{\cal L}=\frac{1}{2}\left[(1+\ktr)\vec{E}^{2}-(1-\ktr)\vec{B}^{2}\right]+\vec{E}\cdot(\kop)\cdot\vec{B}\\+\frac{1}{2}\vec{E}\cdot(\kem)\cdot\vec{E}
+\frac{1}{2}\vec{B}\cdot(\kem)\cdot\vec{B}\;.
\end{multline}
In the boosted frame $S'$, both the fields and the $\tilde{\kappa}$'s transform, while the total Lagrangian remains constant, yielding
\begin{multline}
\!\!\!\!\!
\mathcal{L}=\mathcal{L}'=\frac{1}{2}\left[(1+\ktr')\vec{E'}^{2}-(1-\ktr')\vec{B'}^{2}\right]+\vec{E'}\cdot(\kop')\cdot\vec{B'}\\+\frac{1}{2}\vec{E'}\cdot(\kem')\cdot\vec{E'}
+\frac{1}{2}\vec{B'}\cdot(\kem')\cdot\vec{B'}\;.\label{eq:lageqlagp}
\end{multline}
Since the fields transform normally~\cite{Kostelecky:2002}, the boosted fields $\vec{E'}$ and $\vec{B'}$ can be written in terms of the unprimed fields as~\cite{Jackson:1999}
\begin{align}
\vec{E'}&=\gamma(\vec{E}+\vec{\beta}\times\vec{B})-\frac{\gamma^{2}}{\gamma+1}\vec{\beta}(\vec{\beta}\cdot\vec{E}),\label{eq:Ep}\\
\vec{B'}&=\gamma(\vec{B}-\vec{\beta}\times\vec{E})-\frac{\gamma^{2}}{\gamma+1}\vec{\beta}(\vec{\beta}\cdot\vec{B}).\label{eq:Bp}
\end{align}
Substituting \eqref{eq:Ep} and \eqref{eq:Bp} into \eqref{eq:lageqlagp} allows us to determine the relationship between the primed $\tilde{\kappa}$'s in $S'$ and the unprimed $\tilde{\kappa}$'s in $S$ via the term-by-term equality of all factors of $E_{j}E_{k}$, $B_{j}B_{k}$, and $E_{j}B_{k}$ which appear on both sides.  This yields the following general form of the non-birefringent $\tilde{\kappa}$'s in the boosted frame:
\begin{widetext}
\begin{align}\begin{split}
\ktr'&=\left(1+\frac{|\beta|^{2}}{3}\right)\gamma^{2}\ktr+\frac{2}{3}\gamma^{2}\left[(\bx^{2}-\by^{2})\kemyy+(\bx^{2}-\bz^{2})\kemzz\right]\\
&\quad\quad-\frac{4}{3}\gamma^{2}\left[\bx\by\kemxyp+\bx\bz\kemxzp+\by\bz\kemyzp-\bz\kopxyp+\by\kopxzp-\bx\kopyzp\right],\label{eq:ktrp}
\end{split}\end{align}
\begin{align}\begin{split}
\kemyyp&=\frac{2}{3}\left[(1-3\by^{2})\gamma^{2}-1\right]\ktr+\frac{(\bx^{2}-\bz^{2})\gamma^{2}}{3(\gamma+1)^{2}}\left[1+\gamma(2-\gamma(3\by^{2}-1))\right]\kemzz\\
&\quad+\left[\frac{1}{3}(2+(1-\bz^{2})\gamma^{2})+\frac{\by^{2}(\by^{2}-\bx^{2})\gamma^{4}}{(\gamma+1)^{2}}-\frac{2\by^{2}\gamma^{2}(\gamma-2)}{3(\gamma+1)}\right]\kemyy\\
&\quad+\frac{2\gamma^{2}(2+\gamma+(3\by^{2}-1)\gamma^{2})}{3(\gamma+1)^{2}}\left[\bx\by\kemxy+\bx\bz\kemxz+\by\bz\kemyz\right]-\frac{2\bx\bz\gamma^{2}}{\gamma+1}\kemxz\\
&\quad+\frac{2}{3}\gamma^{2}\left(1-\frac{3\by^{2}\gamma}{\gamma+1}\right)\left[\bz\kopxy-\by\kopxz+\bx\kopyz\right]+2\by\gamma\kopxz,\label{eq:kemp22}
\end{split}\end{align}
\begin{align}\begin{split}
\kemzzp&=\frac{2}{3}\left[(1-3\bz^{2})\gamma^{2}-1\right]\ktr+\frac{(\bx^{2}-\by^{2})\gamma^{2}}{3(\gamma+1)^{2}}\left[1+\gamma(2-\gamma(3\bz^{2}-1))\right]\kemyy\\
&\quad+\left[\frac{1}{3}(2+(1-\by^{2})\gamma^{2})+\frac{\bz^{2}(\bz^{2}-\bx^{2})\gamma^{4}}{(\gamma+1)^{2}}-\frac{2\bz^{2}\gamma^{2}(\gamma-2)}{3(\gamma+1)}\right]\kemzz\\
&\quad+\frac{2\gamma^{2}(2+\gamma+(3\bz^{2}-1)\gamma^{2})}{3(\gamma+1)^{2}}\left[\bx\by\kemxy+\bx\bz\kemxz+\by\bz\kemyz\right]-\frac{2\bx\by\gamma^{2}}{\gamma+1}\kemxy\\
&\quad+\frac{2}{3}\gamma^{2}\left(1-\frac{3\bz^{2}\gamma}{\gamma+1}\right)\left(\bz\kopxy-\by\kopxz+\bz\kopyz\right)-2\bz\gamma\kopxy,\label{eq:kemp33}
\end{split}\end{align}
\begin{align}\begin{split}
\kemxyp&=-2\bx\by\gamma^{2}\ktr+\frac{\bx\by(\by^{2}-\bx^{2})\gamma^{4}}{(\gamma+1)^{2}}\kemyy\\
&\quad-\frac{\bx\by\gamma^{2}(1+\gamma+(\bx^{2}-\bz^{2})\gamma^{2})}{(\gamma+1)^{2}}\kemzz\\
&\quad+\left(1+\frac{\gamma^{2}(\bx^{2}+\by^{2})}{\gamma+1}+\frac{2\bx^{2}\by^{2}\gamma^{4}}{(\gamma+1)^{2}}\right)\kemxy+\frac{\by\bz\gamma^{2}}{\gamma+1}\left(1+\frac{2\bx^{2}\gamma^{2}}{\gamma+1}\right)\kemxz\\
&\quad+\frac{\bx\bz\gamma^{2}}{\gamma+1}\left(1+\frac{2\by^{2}\gamma^{2}}{\gamma+1}\right)\kemyz-\frac{2\bx\by\bz\gamma^{3}}{\gamma+1}\kopxy\\
&\quad+\bx\gamma\left(1+\frac{2\by^{2}\gamma^{2}}{\gamma+1}\right)\kopxz-\by\gamma\left(1+\frac{2\bx^{2}\gamma^{2}}{\gamma+1}\right)\kopyz,\label{eq:kemp12}
\end{split}\end{align}
\begin{align}\begin{split}
\kemxzp&=-2\bx\bz\gamma^{2}\ktr+\frac{\bx\bz(\bz^{2}-\bx^{2})\gamma^{4}}{(\gamma+1)^{2}}\kemzz\\
&\quad-\frac{\bx\bz\gamma^{2}(1+\gamma+(\bx^{2}-\by^{2})\gamma^{2})}{(\gamma+1)^{2}}\kemyy\\
&\quad+\frac{\by\bz\gamma^{2}}{\gamma+1}\left(1+\frac{2\bx^{2}\gamma^{2}}{\gamma+1}\right)\kemxy+\left(1+\frac{\gamma^{2}(\bx^{2}+\bz^{2})}{\gamma+1}+\frac{2\bx^{2}\bz^{2}\gamma^{4}}{(\gamma+1)^{2}}\right)\kemxz\\
&\quad+\frac{\bx\by\gamma^{2}}{\gamma+1}\left(1+\frac{2\bz^{2}\gamma^{2}}{\gamma+1}\right)\kemyz-\bx\gamma\left(1+\frac{2\bz^{2}\gamma^{2}}{\gamma+1}\right)\kopxy\\
&\quad+\frac{2\bx\by\bz\gamma^{3}}{\gamma+1}\kopxz-\bz\gamma\left(1+\frac{2\bx^{2}\gamma^{2}}{\gamma+1}\right)\kopyz,\label{eq:kemp13}
\end{split}\end{align}
\begin{align}\begin{split}
\kemyzp&=-2\by\bz\gamma^{2}\ktr+\frac{\by\bz\gamma^{2}}{\gamma+1}\left(1+\frac{(\by^{2}-\bx^{2})\gamma^{2}}{(\gamma+1)^{2}}\right)\kemyy\\
&\quad+\frac{\by\bz\gamma^{2}}{\gamma+1}\left(1+\frac{(\bz^{2}-\bx^{2})\gamma^{2}}{(\gamma+1)^{2}}\right)\kemzz\\
&\quad+\frac{\bx\bz\gamma^{2}}{\gamma+1}\left(1+\frac{2\by^{2}\gamma^{2}}{\gamma+1}\right)\kemxy+\frac{\bx\by\gamma^{2}}{\gamma+1}\left(1+\frac{2\bz^{2}\gamma^{2}}{\gamma+1}\right)\kemxz\\
&\quad+\left(1+\frac{\gamma^{2}(\by^{2}+\bz^{2})}{\gamma+1}+\frac{2\by^{2}\bz^{2}\gamma^{4}}{(\gamma+1)^{2}}\right)\kemyz-\by\gamma\left(1+\frac{2\bz^{2}\gamma^{2}}{\gamma+1}\right)\kopxy\\
&\quad+\bz\gamma\left(1+\frac{2\by^{2}\gamma^{2}}{\gamma+1}\right)\kopxz-\frac{2\bx\by\bz\gamma^{3}}{\gamma+1}\kopyz,\label{eq:kemp23}
\end{split}\end{align}
\begin{align}\begin{split}
\kopxyp&=2\bz\gamma^{2}\ktr-\frac{(\by^{2}-\bx^{2})\bz\gamma^{3}}{\gamma+1}\kemyy-\left(\bz\gamma+\frac{(\bz^{2}-\bx^{2})\bz\gamma^{3}}{\gamma+1}\right)\kemzz\\
&\quad-\frac{2\bx\by\bz\gamma^{3}}{\gamma+1}\kemxy-\gamma\left(1+\frac{2\bz^{2}\gamma^{2}}{\gamma+1}\right)\left[\bx\kemxz+\by\kemyz\right]\\
&\quad+\gamma\kopxy+\frac{\bz\gamma^{2}(1+2\gamma)}{\gamma+1}\left[\bz\kopxy-\by\kopxz+\bx\kopyz\right],\label{eq:kopp12}
\end{split}\end{align}
\begin{align}\begin{split}
\kopxzp&=-2\by\gamma^{2}\ktr-\left(\by\gamma+\frac{(\by^{2}-\bx^{2})\by\gamma^{3}}{\gamma+1}\right)\kemyy-\frac{(\bz^{2}-\bx^{2})\by\gamma^{3}}{\gamma+1}\kemzz\\
&\quad+\frac{2\bx\by\bz\gamma^{3}}{\gamma+1}\kemxz+\gamma\left(1+\frac{2\by^{2}\gamma^{2}}{\gamma+1}\right)\left[\bx\kemxy+\bz\kemyz\right]\\
&\quad+\gamma\kopxz+\frac{\by\gamma^{2}(1+2\gamma)}{\gamma+1}\left[\bz\kopxy-\by\kopxz+\bx\kopyz\right],\label{eq:kopp13}
\end{split}\end{align}
\begin{align}\begin{split}
\kopyzp&=2\bx\gamma^{2}\ktr+\bx\gamma\left[1-\frac{(\by^{2}-\bx^{2})\gamma^{2}}{\gamma+1}\right]\kemyy+\bx\gamma\left[1-\frac{(\bz^{2}-\bx^{2})\gamma^{2}}{\gamma+1}\right]\kemzz\\
&\quad-\gamma\left(1+\frac{2\bx^{2}\gamma^{2}}{\gamma+1}\right)\left[\by\kemxy+\bz\kemxz\right]-\frac{2\bx\by\bz\gamma^{3}}{\gamma+1}\kemyz\\
&\quad+\gamma\kopyz-\frac{\bx\gamma^{2}(1+2\gamma)}{\gamma+1}\left[\bz\kopxy-\by\kopxz+\bx\kopyz\right].\label{eq:kopp23}
\end{split}\end{align}

\end{widetext}

In the limit that only $\ktr$ has significant value in the SCCEF, the Michelson-Morley observable \eqref{eq:michelsonmorleysens} becomes
\begin{equation}
\frac{\delta\nu}{\nu}=S_{e}(2\gamma^{2})\left\{\left[\by^{2}-\bx^{2}\right]\cos{2\theta}+2\bx\by\sin{2\theta}\right\}(\ktr)_{\odot},
\end{equation}
where $(\ktr)_{\odot}$ represents the value of $\ktr$ in the SCCEF, yielding the same result as reported in \eqref{eq:secondordermmktrdep}, and derived in Appendix \ref{app:relveladd}.

Note that these transformations are only appropriate between concordant frames~\cite{Kostelecky:2001a}, in which the effects of the $\tilde{\kappa}$ parameters are perturbative.  As can be seen from Eqs. \eqref{eq:ktrp} through \eqref{eq:kopp23}, boosts between frames with large relative $\gamma$ can enhance the effects of various $\tilde{\kappa}$'s by up to $\gamma^{2}$ in one frame relative to the other.  In general, $\gamma^{2}$ should be much smaller than the smallest inverse fractional shift $1/\rho$ in the speed of light that is generated by the $\tilde{\kappa}$'s.  Thus the maximum boost relative to the SCCEF for which the above relations are useful in the absence of more complete knowledge of the underlying physics at high energy scale is limited by the most poorly bounded of the $\tilde{\kappa}$'s.

\end{document}